\pdfoutput=1
\documentclass[a4paper]{esannV2}
\usepackage[T1]{fontenc}
\usepackage{lmodern}  %
\usepackage[utf8]{inputenc}
\usepackage{graphicx}
\usepackage{amssymb,amsmath,array}
\usepackage{booktabs,multirow}
\usepackage{url}
\usepackage[bottom]{footmisc}  %
\newcommand{\numkfiveSpeedup}{89}
\newcommand{\numkfiveFgTime}{90}
\newcommand{\numkfiveFgMem}{39}
\newcommand{\numkfiveAcfTime}{1.0}
\newcommand{\numksixAcfTime}{3.5}
\newcommand{\numksixAcfMem}{1.25}
\newcommand{\numksixFgMem}{3.1}
\newcommand{\numksixAcfErr}{$-$0.038\%}

\newcommand{\numkfiveRank}{16}
\newcommand{\numacfFgMaxDiff}{5.1\times10^{-6}}

\newcommand{\numroundCorrTenErrE}{$-$27.1\%}
\newcommand{\numisCorrTenSd}{0.41\%}

\newcommand{\numpropTimeMin}{0.6}
\newcommand{\numpropTimeMax}{3.5}
\newcommand{\numpropSpaMax}{0.001\%}
\newcommand{\numpropEstMax}{1.7\times10^{-5}}
\newcommand{\numpropChi}{19.7}
\newcommand{\numpropDof}{18}
\newcommand{\numpropZmax}{1.8}
\newcommand{\numovSpaTime}{0.2}
\newcommand{\numovAcfTime}{1.4}
\newcommand{\numovFgTime}{90}
\newcommand{\numovIsBand}{28}
\newcommand{\numovIsDomBand}{19}
\newcommand{\numovMcBandDays}{46}
\newcommand{\numovNtwoMcBandDays}{12}

\newcommand{\numovNtwoSpaErr}{0.079\%}
\newcommand{\numovNtwoAcfErr}{0.0093\%}

\newcommand{\numovNtwoIsBand}{105}
\newcommand{\numovNtwoIsDomBand}{10}
\newcommand{\numovVarRatioThou}{1.5}
\newcommand{\numovVarRatioTwo}{11}
\newcommand{\numovDomSpeedThou}{13}
\newcommand{\numovDomSpeedTwo}{5}
\newcommand{\numovEpsStop}{10^{-5}}
\newcommand{\numovLadEpsHi}{10^{-4}}
\newcommand{\numovLadEpsLo}{10^{-8}}
\newcommand{\numovLadTimeMinThou}{0.5}
\newcommand{\numovLadTimeMaxThou}{6.3}
\newcommand{\numovLadTimeMinTwo}{0.5}
\newcommand{\numovLadTimeMaxTwo}{25}
\newcommand{\numsetMax}{4.1\times10^{-5}}
\newcommand{\numfdCases}{22}
\newcommand{\numfdWrong}{12}

\newcommand{\numfdSpaMax}{6\%}
\newcommand{\numeNineN}{442}
\newcommand{\numeNineVars}{10}
\newcommand{\numeNineScreened}{45}
\newcommand{\numeNinePairs}{15}

\newcommand{\numeNineChi}{14.5}

\newcommand{\numeNineTimeMax}{5.0}

\newcommand{\numeNineMcMaxDays}{7}

\newcommand{\numeNineIsMax}{0.8}
\newcommand{\numeNineSpaMax}{0.02\%}
\newcommand{\numeNineKurtMax}{1.0}

\newcommand{\R}{\mathbb{R}}
\newcommand{\E}{\mathbb{E}}
\newcommand{\vf}{\boldsymbol{f}}
\newcommand{\vs}{\boldsymbol{s}}
\newcommand{\vS}{\boldsymbol{S}}
\newcommand{\vU}{\boldsymbol{U}}

\newcommand{\vom}{\boldsymbol{\omega}}
\newcommand{\vlam}{\boldsymbol{\lambda}}
\newcommand{\vbeta}{\boldsymbol{\beta}}
\newcommand{\vmu}{\boldsymbol{\mu}}
\newcommand{\acf}{\mbox{TT-Bootstrap}}  %

\begin{document}

\title{Bootstrap p-values without Monte Carlo error\\ for light-tailed data via tilted tensor trains}

\author{Hiroki Hasegawa%
\thanks{This work was supported by JST SPRING, Grant Number JPMJSP2124. This research was
partially supported by DSEP at University of Tsukuba. Claude (Anthropic) assisted with
English proofreading and with preparing the figures and tables.}
\vspace{.3cm}\\
Graduate School of Science and Technology, University of Tsukuba\\
Tsukuba -- Japan\\
}

\maketitle

\begin{abstract}
Machine learning and genomics select features with bootstrap and
permutation tests, so multiple-testing thresholds fall to $10^{-7}$ and below. Plain Monte Carlo needs
billions of resamples per test. We propose a method for the ideal bootstrap p-value
$\mathrm{P}^*(T^*\ge t_{\mathrm{obs}})$ without Monte Carlo error. For smooth statistics of weighted sums, we compress the sums' tilted
characteristic function into a tensor train, then invert and
integrate it. At $n=1000$, Gaussian-data p-values down to $10^{-10}$ took
\mbox{\numpropTimeMin--\numpropTimeMax\,s}, consistent with 0.1\%-accurate references.
For real-data correlation tests, plain Monte Carlo would need up to \numeNineMcMaxDays{} days for
1\% relative error, whereas ours took at most \numeNineTimeMax\,s.
\end{abstract}

\section{Introduction}

\begin{figure}[b]
\centering
\includegraphics[width=\textwidth]{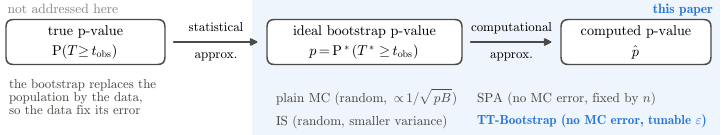}
\caption{Two approximations in a bootstrap p-value (IS = importance sampling, SPA =
saddlepoint approximation). We address only the computational one.}\label{fig:scope}
\end{figure}

Machine learning and genomics select features by resampling tests
such as permutation tests~\cite{Francois2006,Che2014}. Testing all pairwise correlations
of 1000 features means $5\times10^5$ tests, so the Bonferroni threshold is $10^{-7}$.
Genome-wide association studies use $5\times10^{-8}$, the Bonferroni threshold for
$10^6$ tests. There, a permutation p-value needs at least $2\times10^7$ resamples per
test to control the type I error rate, and more to estimate it accurately~\cite{Che2014}.
The bootstrap can be costly for large data~\cite{Kleiner2014}. Repeated over many tests and
studies worldwide, computing bootstrap p-values can add up to a large energy and carbon
footprint~\cite{Lannelongue2021}.

A bootstrap p-value involves two approximations~\cite{Efron1979,DavisonHinkley1997}
(Fig.~\ref{fig:scope}). First, the bootstrap approximates the unknown null distribution of a test
statistic $T$ by resampling the data under the null. Given the data, the statistic and the
resampling scheme, this step defines the ideal bootstrap p-value
$p=\mathrm{P}^*(T^*\ge t_{\mathrm{obs}})$. Second, $p$ is a sum over astronomically many
resamples, so plain Monte Carlo (MC) estimates it from $B$ of them, with relative error
$\approx1/\sqrt{pB}$. This paper addresses only the second approximation. Tilted
importance sampling (IS)~\cite{Johns1988,Hesterberg1995} reduces the variance, but it remains
random, and its efficiency depends on the tilt. Saddlepoint approximations
(SPA)~\cite{LugannaniRice1980,DavisonHinkley1988,JingRobinson1994} are fast and accurate for
large $n$, but more computation cannot reduce their error at a given $n$.

Many statistics are smooth functions $T=g(\vS)$ of $k$ weighted sums
$\vS=\sum_i w_i\vf_i\in\R^k$ with bootstrap weights $w_i$. The joint characteristic function (CF) of
$\vS$ has a closed form, so Fourier inversion can compute the bootstrap
distribution. In 1D, methods without MC error include shifted
FFTs~\cite{KeichNagarajan2006},
exact algorithms for the mean~\cite{BertsimasSturt2020} and quantized tensor trains
(TT)~\cite{RodriguezAldavero2026}. For $k\ge2$, the frequency grid
grows exponentially with $k$. TT-compressed CFs, with fixed
damping~\cite{KastoryanoPancotti2022} or none~\cite{ArensteinKastoryano2026}, avoid this
growth for multi-asset options, and damping can be tuned for lower error~\cite{Bayer2023}.
These TT methods compute a different quantity, an expected payoff over a 1D or box-shaped region. Bootstrap p-values are tiny probabilities of a nonlinear region and need a small
relative error.

We propose \acf{}, which computes the ideal $p$ of the usual bootstrap without MC error in
three steps. First, it
tilts the weights (i.e., damping) so that the tail of interest becomes the bulk, because TT rounding controls
only absolute errors. Second, it represents the tilted CF as
a TT (a matrix product state) by TT-cross~\cite{OseledetsTyrtyshnikov2010}. Third, it integrates
over the region $\{g(\vs)\ge t\}$ with 1D inverse transforms per TT core and an
exact integral along the last axis. For Gaussian data with $n=1000$, it computed p-values
down to $10^{-10}$ in \numpropTimeMin--\numpropTimeMax\,s.

\section{Problem setting}

Let $\vf_i\in\R^k$, $i=1,\dots,n$, be features of the data. The bootstrap weights
$\boldsymbol{w}$ are multinomial, $\boldsymbol{w}\sim\mathrm{Mult}(n;1/n,\dots,1/n)$, or
independent Poisson$(1)$. We seek $p=\mathrm{P}^*\big(g(\vS)\ge t\big)$ with
$\vS=\sum_{i=1}^n w_i\vf_i$. For a test of zero correlation between $x$ and $y$, we first
impose the null~\cite{DavisonHinkley1997} by setting $\tilde y_i=y_i-\hat b x_i$
with the least-squares slope $\hat b$. We then use $k=5$ features
$\vf_i=(x_i,\,\tilde y_i,\,x_i^2,\,\tilde y_i^2,\,x_i\tilde y_i)$. With Poisson weights, the
total weight is random, so $\vf_i$ also contains $1$ ($k=6$). The regression slope uses
$(x_i,\tilde y_i,x_i^2,x_i\tilde y_i)$ ($k=4$). Here $k$ is the number of sums in one
statistic, not the dimension of the data. In all cases, $g$ increases in the last
sum with the others fixed, so $\{g(\vs)\ge t\}=\{s_k\ge\tau(s_{1:k-1})\}$ with an explicit
$\tau$.

The exponential tilt of the weights has a closed form. For $\vlam\in\R^k$, let $Q_{\vlam}$ be
the multinomial law with cell probabilities $\pi_i\propto e^{\vlam\cdot\vf_i}$, or the Poisson
law with means $e^{\vlam\cdot\vf_i}$. Then $dQ_{\vlam}/dP=\exp(\vlam\cdot\vS-K(\vlam))$ with
$K(\vlam)=\log\E e^{\vlam\cdot\vS}$, and the CF of $\vS$ under $Q_{\vlam}$ is
\begin{equation*}
\varphi_{\vlam}(\vom)=\Big(\sum_i \pi_i e^{\mathrm{i}\vom\cdot\vf_i}\Big)^{n}
\quad\text{or}\quad
\varphi_{\vlam}(\vom)=\exp\Big(\sum_i e^{\vlam\cdot\vf_i}\big(e^{\mathrm{i}\vom\cdot\vf_i}-1\big)\Big).
\end{equation*}

\section{Method}\label{sec:method}

\textbf{Tilting and change of variables.} To keep the relative error small, we tilt the weights so that
the region carries much of the mass. Let $\vmu_{\vlam}$ and $\Sigma_{\vlam}=LL^\top$ be the
mean and covariance of $\vS$ under $Q_{\vlam}$, with $L$ lower triangular. The new variable
$\vU=L^{-1}(\vS-\vmu_{\vlam})$ has mean zero and identity covariance. Changing the measure
gives the exact identity
\begin{equation}\label{eq:tilt}
p=e^{K(\vlam)-\vlam\cdot\vmu_{\vlam}}\;\E_{Q_{\vlam}}\!\big[\mathbf{1}\{g(\vS)\ge t\}\,
e^{-\vbeta\cdot\vU}\big],\qquad \vbeta=L^\top\vlam .
\end{equation}
We take $\vlam$ along $\nabla g$ at the untilted mean, scaled so that $g(\vmu_{\vlam})=t$
(the gradient tilt). This centers the tilted law on a boundary point, a cheap proxy for the dominating
point~\cite{SadowskyBucklew1990}. As $L$ is lower triangular, the region becomes
$u_k\ge h(u_{1:k-1})$.

\textbf{TT representation of the CF.} To avoid the exponential growth of the frequency grid,
we compress the CF into a TT. We restrict each $U_j$ to a window $[-W_j,W_j)$ with tail mass
below $10^{-12}$. We expand the density of $\vU$ in a Fourier series with
period $P_j=2W_j$ and frequencies $\omega_{j,m}=2\pi m/P_j$, $|m|\le M_j$, where the CF along
axis $j$ falls below $10^{-18}$ for $|m|>M_j$. The coefficients
$\Phi[\boldsymbol{m}]=\varphi_U(\omega_{1,m_1},\dots,\omega_{k,m_k})$ form a $k$-way tensor
with $\prod_j(2M_j+1)\approx 73^k$ entries ($2\times10^9$ for $k=5$). We approximate it as a
TT, $\Phi[\boldsymbol{m}]\approx G_1[m_1]\cdots G_k[m_k]$ with
$G_j[m_j]\in\mathbb{C}^{r_{j-1}\times r_j}$~\cite{Oseledets2011}.
TT-cross~\cite{OseledetsTyrtyshnikov2010} builds this TT from only $O(kr^2\max_j M_j)$
CF evaluations ($O(nk)$ each), and we then round it to a relative tolerance $\varepsilon$.

\textbf{Integration over the region.} To obtain $p$, we insert the Fourier series
into~\eqref{eq:tilt} and integrate in TT format. We integrate the head coordinates
$u_1,\dots,u_{k-1}$ on grids of spacing $0.6$, and the last one exactly,
\begin{equation*}
\int_{h}^{W_k}\! e^{-\mathrm{i}\omega_{k,m}u-\beta_k u}\,\frac{du}{P_k}
=\frac{e^{-z_m h}-e^{-z_m W_k}}{z_m P_k},\qquad z_m=\beta_k+\mathrm{i}\,\omega_{k,m}.
\end{equation*}
A small matrix with $e^{-\beta_j u_j}$ maps each head core to the $u_j$ grid. We contract the last core
with the vector above, which depends on $h(u_{1:k-1})$, and evaluate the head-grid sum by
blockwise contraction. The CF is thus never stored in full.

\textbf{Adaptive tolerance.} The $\varepsilon$ needed for a given relative error is not known
in advance. We therefore build the TT for $\varepsilon=10^{-4},\dots,10^{-8}$ and stop when
two consecutive estimates differ by at most $10^{-3}$ (relative), or else we flag a failure. The difference serves
as an error estimate, and times include all builds. All runs use double
precision in PyTorch on one GPU (NVIDIA A100, 40\,GB).

\section{Experiments}

\textbf{Setup.} Data are bivariate Gaussian ($n\in\{50,100,200,1000\}$, correlation $0.3$) with
the null imposed and multinomial weights unless noted. We set $t$ so that the SPA gives
$p=10^{-2},\dots,10^{-10}$. We compare three methods without MC error (\acf{}, the
SPA~\cite{JingRobinson1994}, full-grid shifted FFTs~\cite{KeichNagarajan2006}) with two
sampling ones, plain MC and tilted IS. As in~\cite{Hesterberg1995}, tilted IS draws 10\% of
weights from $P$, 90\% from $Q_{\vlam}$ (mixture $R$, so $dP/dR\le10$) and averages ${\mathbf{1}\{g(\vS)\ge t\}\,dP/dR}$
with the gradient tilt or the dominating point (the likeliest point of the region, known to
make IS asymptotically efficient~\cite{SadowskyBucklew1990}). References (proxies for the
exact $p$) are IS runs to a relative standard error (s.e.) $\le10^{-3}$. Times include setup.

\begin{figure}[t]
\centering
\includegraphics[width=\textwidth]{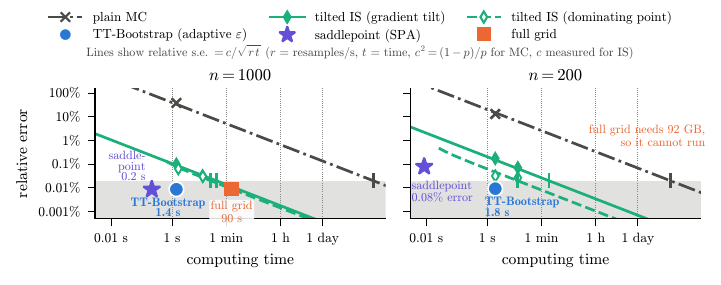}
\caption{Relative error against computing time (log--log), correlation test ($k=5$,
$p\approx10^{-6}$). All methods ran in one A100 session (SPA on the CPU). The gray
band marks errors within 2 s.e.\ of the reference (s.e.\ $10^{-4}$). Ticks mark where
sampling lines reach the band (extrapolated from 10 runs at two budgets). Tilted IS reaches it after
\numovIsDomBand\,s (dominating point) and \numovIsBand\,s (gradient tilt) at $n=1000$ and after
\numovNtwoIsDomBand\,s and \numovNtwoIsBand\,s at $n=200$, plain MC after \numovMcBandDays{} and
\numovNtwoMcBandDays{} days.}\label{fig:overview}
\end{figure}

\begin{figure}[!b]
\centering
\includegraphics[width=\textwidth]{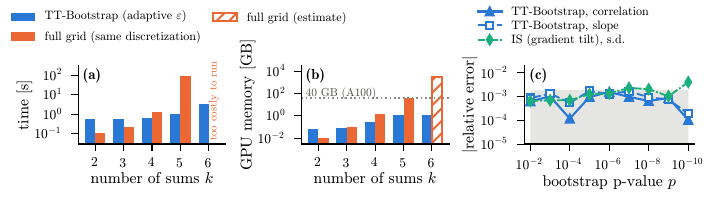}
\caption{(a) Time and (b) GPU memory against $k$ ($n=1000$, $p\approx10^{-6}$). (c) Relative
error against $p$ ($n=1000$). Gray shows errors below 2 s.e.\ of the
reference. IS s.d.\ uses 10 runs as long as ours.}\label{fig:k}
\end{figure}

\textbf{Error and cost against all baselines (Fig.~\ref{fig:overview}).} MC and IS errors fall
as $\smash{B^{-1/2}}$ with the number of resamples $B\propto\text{time}$, giving lines of slope $-1/2$
on these log--log axes. Errors in the band are within the reference's own uncertainty.
At $n=1000$ (left), the SPA (\numovSpaTime\,s), \acf{} (\numovAcfTime\,s) and the full
grid (\numovFgTime\,s) fall in the band and differ only in time. The adaptive rule stopped at $\varepsilon=\numovEpsStop$, and every fixed $\varepsilon$
from $\numovLadEpsHi$ to $\numovLadEpsLo$ also fell in the band
(\numovLadTimeMinThou--\numovLadTimeMaxThou\,s at $n=1000$,
\numovLadTimeMinTwo--\numovLadTimeMaxTwo\,s at $n=200$). At $n=200$ (right), \acf{} stays in the band
(\numovNtwoAcfErr{}), but the SPA is above it (\numovNtwoSpaErr{}). The
dominating point made IS \numovVarRatioTwo$\times$ more efficient, but IS would still need
\numovNtwoIsDomBand\,s, \numovDomSpeedTwo$\times$ our time (\numovDomSpeedThou$\times$
at $n=1000$).

\textbf{Why tensor trains (Fig.~\ref{fig:k}a, b).} We used the variance, covariance, slope,
correlation and Poisson-weight correlation ($k=2,\dots,6$). With equal tilt and discretization, the
full grid is faster than \acf{} for $k\le3$. However, its grid grows as
$\approx73^k$. At $k=5$, it used \numkfiveFgTime\,s and \numkfiveFgMem\,GB,
\numkfiveSpeedup$\times$ longer than ours (\numkfiveAcfTime\,s, TT rank \numkfiveRank). At
$k=6$, it would need \numksixFgMem\,TB, whereas ours used
\numksixAcfTime\,s and \numksixAcfMem\,GB (error \numksixAcfErr{}). Where both ran,
they agreed within $\numacfFgMaxDiff$.

\textbf{Very small p-values (Fig.~\ref{fig:k}c).} From $p=10^{-2}$ to $10^{-10}$, \acf{}
agreed with the reference for the slope and correlation (within \numpropZmax{} s.e.,
$\chi^2=\numpropChi$ over \numpropDof{} cases). The adaptive rule stopped at $\varepsilon=10^{-5}$ for all $p$
(\numpropTimeMin--\numpropTimeMax\,s). Its error estimate was up to $11\times$ too small,
but the difference from the $\varepsilon=10^{-8}$ result never exceeded $\numpropEstMax$.
Without tilting, the error at $p=10^{-10}$ was \numroundCorrTenErrE{} even at $\varepsilon=10^{-8}$.
In the same time, IS had an s.d.\ of \numisCorrTenSd{} at $p=10^{-10}$. The SPA agreed
with ours within \numpropSpaMax{}. Changing the TT-cross seed, grid spacing or
window and frequency tolerances moved $p$ by $\le\numsetMax$ (relative).

\textbf{Real data (Table~\ref{tab:e9}).} To check \acf{} on real data, we ran the
zero-correlation test of Section~2 ($t=$ observed $|r|$) on all
\numeNineScreened{} pairs of \numeNineVars{} continuous variables of the diabetes
data~\cite{Efron2004} (untransformed, excess kurtosis $\le\numeNineKurtMax$).
For the \numeNinePairs{} pairs the SPA put in $10^{-10}\le p\le10^{-4}$ (25 were smaller), \acf{} matched the
references ($\chi^2=\numeNineChi$). Yet dominating-point IS
(0.1\% s.e.\ in \numeNineIsMax\,s) and the SPA (within \numeNineSpaMax{} of ours) were
also fast and accurate.

\begin{table}[!h]
\centering
\caption{Diabetes data ($n=\numeNineN$, BP = blood pressure, TC and LDL = total and LDL cholesterol,
LTG = log triglycerides, Y = progression), 5 of the
\numeNinePairs{} tests with $10^{-10}\le p\le10^{-4}$ ($r$ = observed correlation, $p$ = IS
reference~\cite{Hesterberg1995}, against which errors are measured). $^\dagger$Plain MC run as long as ours,
$\pm$ its s.e. `Time to 1\%' is its time to a 1\% s.e.}\label{tab:e9}
{\footnotesize\setlength{\tabcolsep}{3.5pt}
\begin{tabular}{lrrrrrr}
\toprule
 & & & \multicolumn{2}{c}{\acf{} (ours)} & \multicolumn{2}{c}{plain MC~\cite{Efron1979}} \\
\cmidrule(lr){4-5}\cmidrule(lr){6-7}
pair & $r$ & $p$ & error & time & error$^\dagger$ & time to 1\% \\
\midrule
BP--LDL & $+0.19$ & $3.1\times10^{-5}$ & +0.08\% & 1.6\,s & $-$0.8$\pm$4.0\% & 25\,s \\
TC--Y & $+0.21$ & $9.4\times10^{-7}$ & +0.08\% & 4.7\,s & $-$3.1$\pm$13\% & 14\,min \\
BP--TC & $+0.24$ & $1.0\times10^{-7}$ & +0.20\% & 1.6\,s & $-$5.7$\pm$71\% & 2\,h \\
age--TC & $+0.26$ & $1.5\times10^{-8}$ & $-$0.10\% & 4.7\,s & $-$100\% & 15\,h \\
age--LTG & $+0.27$ & $1.4\times10^{-9}$ & +0.08\% & 3.3\,s & $-$100\% & 7\,d \\
\bottomrule

\end{tabular}}
\end{table}

\section{Discussion and conclusion}\label{sec:disc}

\acf{} was faster than plain MC, more so at smaller $p$, and than the
full grid for $k\ge4$. The SPA was faster but less accurate at $n=200$
(Fig.~\ref{fig:overview}). \acf{} has no MC error and flagged its failures in our stress
tests, so it may reproducibly check SPA or IS p-values near significance thresholds.

Several limitations remain. \acf{} failed for small $n$ and very small $p$, probably
because $e^{-\vbeta\cdot\vU}$ in~\eqref{eq:tilt} amplifies TT errors far from the center. In
\numfdCases{} Gaussian stress cases ($n=50$--$200$, $p=10^{-4}$--$10^{-10}$), its error
exceeded 1\% in \numfdWrong{}. The adaptive rule did not converge in exactly these
\numfdWrong{}, so a user can detect them and switch to the SPA (error $\le$~\numfdSpaMax{}).
The adaptive error estimate, however, is heuristic, not a certified bound. The head-grid
sum has $\approx34^{k-1}$ terms and took 1.5 of the 3.5\,s at $k=6$, so larger $k$ may need
a TT-structured integral. Fig.~\ref{fig:k}c compares only with gradient-tilt IS, and
dominating-point IS had \numovVarRatioThou--\numovVarRatioTwo$\times$ lower variance at
$p\approx10^{-6}$. We tested only Gaussian or light-tailed data, not heavy-tailed data.
Finally, \acf{} computes the ideal bootstrap p-value, not the true one, and at
$10^{-10}$ the two need not be close~\cite{RobinsonSkovgaard1998}.

{\clubpenalty=10000  %
In conclusion, tensor trains made Fourier inversion for bootstrap tail probabilities
practical up to $k=6$ in our tests. With tilting, \acf{} gave p-values without MC error
down to $10^{-10}$ for Gaussian data with $n=1000$.\par}

\begin{footnotesize}

\end{footnotesize}

\end{document}